\documentclass[journal]{IEEEtran}
\usepackage{cite}
\usepackage[pdftex]{graphicx}
  \DeclareGraphicsExtensions{.pdf,.jpeg,.png}
  \graphicspath{{./fig/}}
\usepackage{amsmath}
\usepackage{autobreak}
\usepackage{amsthm}
  \theoremstyle{definition}
  \newtheorem{definition}{Definition}[section]
\usepackage{amssymb}
\usepackage{bbm}
\usepackage[ruled]{algorithm2e}
\usepackage{array}
\usepackage[hidelinks]{hyperref}
\usepackage{cleveref}
\ifCLASSOPTIONcompsoc
  \usepackage[caption=false,font=normalsize,labelfont=sf,textfont=sf]{subfig}
\else
  \usepackage[caption=false,font=footnotesize]{subfig}
\fi
\usepackage{threeparttable}
\usepackage{tabularx}
\usepackage{multirow}
\usepackage{multicol}
\usepackage{booktabs}
\usepackage{url}
\usepackage[x11names]{xcolor}
\usepackage{soul}
  \soulregister\cite7
  \soulregister\cref7
  \soulregister\Cref7
  
\usepackage{tikz}
  \usetikzlibrary{calc}
\usepackage{lipsum}

\makeatletter 
\newcount\SOUL@minus
\makeatother  

\begin{document}
\title{Bi-level Off-policy Reinforcement Learning for Volt/VAR Control Involving Continuous and Discrete Devices}
\author{Haotian~Liu, ~\IEEEmembership{Graduate~Student~Member,~IEEE},
  Wenchuan~Wu,~\IEEEmembership{Fellow,~IEEE}
  \thanks{This work was supported in part by State Grid Corporation of China Project ``Research on Coordinated Strategy of Multi-type Controllable Resources Based on Collective Intelligence in an Energy Internet Environment''.}
  \thanks{H. Liu and W. Wu (Corresponding Author)  are with the State Key Laboratory of Power Systems, Department of Electrical Engineering, Tsinghua University, Beijing 100084, China (email:lht18@mails.tsinghua.edu.cn, wuwench@tsinghua.edu.cn).}
}
\markboth{Journal of \LaTeX\ Class Files,~Vol.~xx, No.~x, August~20xx}%
{Shell \MakeLowercase{\textit{et al.}}: Bare Demo of IEEEtran.cls for IEEE Journals}
\maketitle

\begin{abstract}
  In Volt/Var control (VVC) of active distribution networks(ADNs), both slow timescale discrete devices (STDDs) and fast timescale continuous devices (FTCDs) are involved.
  The STDDs such as on-load tap changers (OLTC) and FTCDs such as distributed generators should be coordinated in time sequence.
  Such VCC is formulated as a two-timescale optimization problem to jointly optimize FTCDs and STDDs in ADNs.
  Traditional optimization methods are heavily based on accurate models of the system, but sometimes impractical because of their unaffordable effort on modelling.
  In this paper, a novel bi-level off-policy reinforcement learning (RL) algorithm is proposed to solve this problem in a model-free manner.
  A Bi-level Markov decision process (BMDP) is defined to describe the two-timescale VVC problem and separate agents are set up for the slow and fast timescale sub-problems.
  For the fast timescale sub-problem, we adopt an off-policy RL method soft actor-critic with high sample efficiency.
  For the slow one, we develop an off-policy multi-discrete soft actor-critic (MDSAC) algorithm to address the curse of dimensionality with various STDDs.
  To mitigate the non-stationary issue existing the two agents' learning processes, we  propose a multi-timescale off-policy correction (MTOPC) method by adopting importance sampling technique.
  Comprehensive numerical studies not only demonstrate that the proposed method can achieve stable and satisfactory optimization of both STDDs and FTCDs without any model information, but also support that the proposed method outperforms existing two-timescale VVC methods.
\end{abstract}

\IEEEpeerreviewmaketitle

\section{Introduction}
\IEEEPARstart{W}{ith} increasing penetration level of distributed generations (DG) \cite{kurbatovaGlobalTrendsRenewable2020}, modern distribution networks are challenged with severe operating problems such as voltage violations and high network losses. As the common practice, active distribution networks (ADN) have integrated Volt/VAR control (VVC) to optimize the voltage profile and reduce network losses by employing not only the discrete regulation equipments such as on-load tap changers (OLTC) and capacitor banks (CB), but also the continuous control facilities such as the DGs and static var compensators (SVC).

Typically the original VVC task is described as a mixed integer nonlinear programming problem with variables standing for strategies of voltage regulation devices and reactive power resources.
While general symbolic solutions of such problem are hardly available, a variety of methods have been studied and led to various schemes of VVC, which could be categorized via control architectures into the centralized VVC \cite{liuReactivePowerVoltage2009,borghettiUsingMixedInteger2013},
distributed VVC \cite{liuDistributedVoltageControl2018,xuAcceleratedADMMBasedFully2020},
decentralized VVC \cite{zhuFastLocalVoltage2016,liuOnlineMultiagentReinforcement2021}.

Even though the existing VVC methods have achieved considerable performance in traditional distribution networks, most of them rely heavily on the accurate network model.
These model-based methods are seriously challenged when an accurate model is expensive or sometimes impractical to maintain in a fast developing ADN with increasing complexity and numerous components \cite{arnoldModelFreeOptimalControl2016,liuOnlineMultiagentReinforcement2021,wangSafeOffpolicyDeep2019,gaoBatchConstrainedReinforcementLearning2020}.
In the recent years, the research on deep reinforcement learning (DRL) has shown desirable potential on coping with the incomplete model challenges in video game \cite{nachumDataEfficientHierarchicalReinforcement2018,ecoffetFirstReturnThen2021,lazaridisDeepReinforcementLearning2020} and multiple areas in power grid operation, including energy trading \cite{serbanArtificialIntelligenceSmart2020}, network reconfiguration \cite{gaoBatchConstrainedReinforcementLearning2020}, frequency control \cite{stanojevReinforcementLearningApproach2020,zhangResearchAGCPerformance2020} and so on.
Hence, many inspiring DRL-based VVC methods has been proposed recently such as \cite{liuTwostageDeepReinforcement2020,wangSafeOffpolicyDeep2019,liCoordinationPVSmart2019,caoMultiAgentDeepReinforcement2020,liuOnlineMultiagentReinforcement2021,yangTwoTimescaleVoltageControl2020}.
Such DRL-based VVC methods have empowered the agent in the ADN operating utility to learn a near-optimal strategy by interacting with the actual ADN and mining the optimization process data without an accurate ADN model \cite{chenReinforcementLearningDecisionMaking}.

\begin{figure}[h]
  \centering
  \includegraphics[width=1\linewidth]{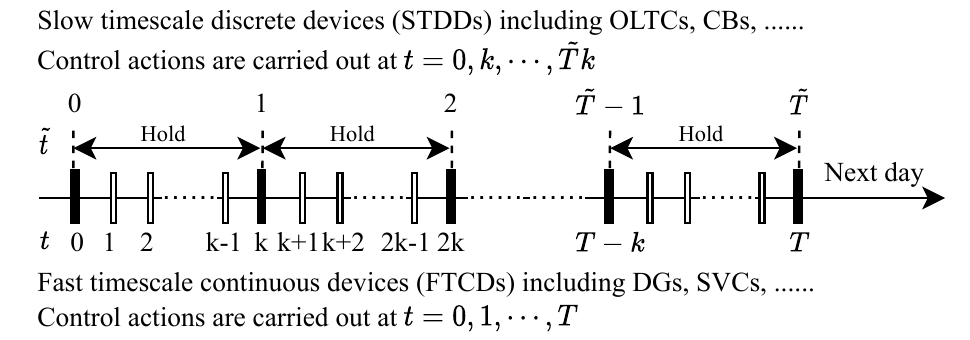}
  \caption{The timeline of two types of devices in the ADNs.}
  \label{fig:timeline}
\end{figure}

Moreover, the characteristics of controlled devices determine the nature of the VVC problem.
Two types of devices considered by modern VVC in ADNs are described in \cref{tbl:devices}, including the Slow Timescale Discrete Devices (STDD) and Fast Timescale Continuous Devices (FTCD).
As shown in \cref{fig:timeline}, we assume FTCDs take $k$ steps in one STDD step.

\begin{table}[h]
  \centering
  \begin{threeparttable}
    \caption{Two Types of Devices Considered by Modern VVC in ADNs}
    \label{tbl:devices}
  \begin{tabular}{lcc}
      \toprule
      Item\hspace{1cm} & \hspace{10mm}STDD\hspace{10mm} & \hspace{10mm}FTCD\hspace{10mm} \\
      \midrule
      Variable & Discrete & Continuous \\
      Timescale & Slow (in hours) & Fast (in minutes)\tnote{o} \\
      Number & Relatively small & Large \\
      Control Price & Limited switching times & Flexible \\
      Devices & OLTCs, CBs, ... & DGs, SVCs\tnote{*}, ... \\
      \bottomrule
  \end{tabular}
  \begin{tablenotes}
    \item[o] The fast timescale depends heavily on communications.
    \item[*] The location of SVCs is similar to STDD. 
  \end{tablenotes}
\end{threeparttable}
\end{table}

In the active distribution networks with both STDDs and FTCDs, all control devices are supposed to work concurrently and cooperatively, while most of the existing VVC works consider either of them.
Because of the huge difference of the natures listed in \cref{tbl:devices} especially the timescale \cref{fig:timeline}, a proper optimization and control method which fully utilizes the fast speed of FTCDs and the limited STDD actions is non-trivial.

With the assumption of oracle models of ADNs, some researchers have developed optimization-based multi-timescale VVC methods such as \cite{zhengRobustReactivePower2017,xuMultiTimescaleCoordinatedVoltage2017,jinTwoTimescaleMultiObjectiveCoordinated2019,jhaBiLevelVoltVAROptimization2019,zafarMultiTimescaleVoltageStabilityConstrained2020,yangTwoTimescaleVoltageControl2020}.
For example, \cite{jinTwoTimescaleMultiObjectiveCoordinated2019} presents a multi-object coordinated VVC method with the slow timescale as MINLP to optimize the power loss and control actions, and the fast timescale as NLP to minimize the voltage deviation.
\cite{jinTwoTimescaleMultiObjectiveCoordinated2019} handles the two stages separately by applying a searching method to solve the MINLP and a scenario-based method to solve the NLP.
Instead, \cite{zhengRobustReactivePower2017} formulates the slow timescale VVC as a robust optimization problem and guarantees the worst case in the fast timescale, which may leads to conservativeness or poor convergence.
To solve the slow timescale MINLP more efficiently online, \cite{yangTwoTimescaleVoltageControl2020} has incorporated a DRL algorithm called deep Q network (DQN) to boost the solution, while optimizing the FTCDs with a model-based second-order cone program.

However, to coordinate the FTCDs and STDDs in two timescales and conduct efficient VVC in a model-free manner, it is indispensable for us to develop DRL-based multi-timescale VVC methods.
Among the DRL-based VVC methods, most of them target at single-timescale.
For example, authors in \cite{wangSafeOffpolicyDeep2019} proposed a safe off-policy RL algorithm to optimize STDDs hourly by formulating the voltage constraints explicitly and considering the device switching cost.
In contrast, references \cite{caoMultiAgentDeepReinforcement2020,liuOnlineMultiagentReinforcement2021} are designed to optimize the FTCDs in minutes by incorporating and improving continuous RL algorithms.
As for multi-timescale (two-timescale) VVC, reference \cite{yangTwoTimescaleVoltageControl2020} applied DQN to the slow timescale optimization problem, but depended on oracle model in the fast timescale optimization.
Research on DRL-based VVC for both fast and slow timescales, which could achieve model-free optimization, is still urgently needed.

\begin{figure}[h]
  \centering
  \includegraphics[width=1\linewidth]{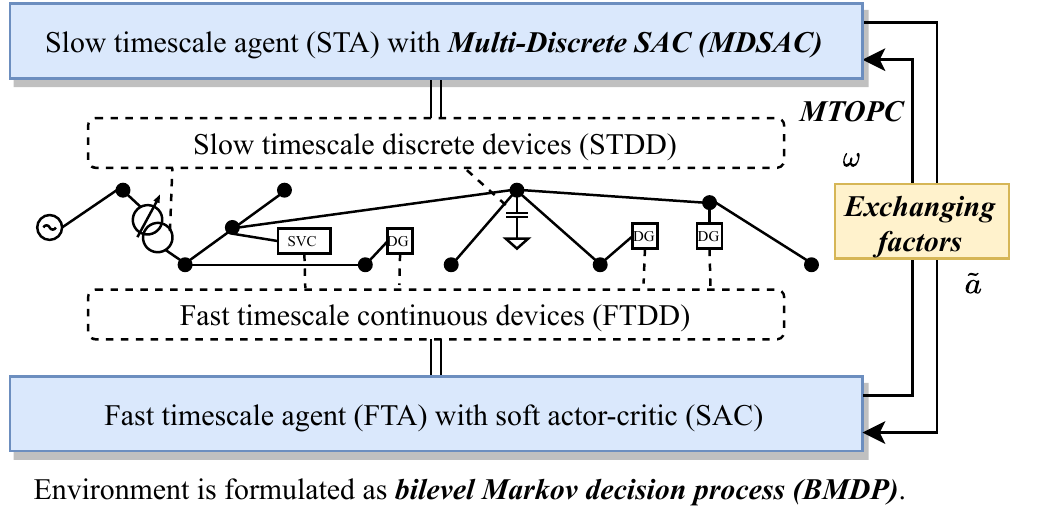}
  \caption{Overall structure of the proposed bi-level off-policy RL for two-timescale VVC in ADNs. The contributions of this paper are \textbf{\textit{highlighted}}.}
  \label{fig:proposed}
\end{figure}

Unfortunately, DRL algorithms for two-timescale agent with such different natures shown in \cref{tbl:devices} are non-trivial and rarely studied.
A reasonable solution is to set up RL agents for both timescales individually like \cref{fig:proposed}, so as to satisfy natures of FTCDs and STDDs.
However, for the training processes of the agents, traditional RL approaches such as Q-learning are poorly suited.
The most fundamental issue is that the policy of the fast timescale agent (FTA) in the lower layer is changing as training processes, and the environment becomes non-stationary from the perspective of slow timescale agent (STA) in the upper layer \cite{loweMultiAgentActorCriticMixed2020}.
In another way, at every decision time of STA, the next step is not only depended on the STDDs' actions of STA itself, but also depended on the subsequent FTCDs' actions of FTA.
This issue extremely challenges the learning stability and prevents the use of experience replay off-policy RL algorithms, which are generally more efficient than the on-policy ones \cite{guQPropSampleEfficientPolicy2017}.
Besides, the STA involves multiple STDDs and is bothered by the curse of dimensionality in action space.

In this paper, we propose a novel bi-level off-policy RL algorithm and develop a two-timescale VVC accordingly to jointly optimize FTCDs and STDDs in ADNs in a model-free manner.
As shown in \cref{fig:proposed}, we first formulate the two-timescale VVC problem in the bi-level RL framework with separate STA and FTA established.
Then, the two agents are implemented with detailed designed actor-critic algorithms.
Finally, STA and FTA are trained jointly by introducing the multi-timescale off-policy correction technique to eliminate the non-stationary problem.
The proposed model-free two-timescale VVC method not only ensures the stability of learning by the coordination of STA and FTA, but also performs off-policy learning with desirable sample efficiency.

Comparing with previous studies on VVC in ADNs, the unique contributions of this paper are summarized as follows.
\begin{enumerate}
  \item To realize model-free optimization of FTCDs and STDDs described \cref{tbl:devices} together, we design a mathematical formulation called bi-level Markov decision process to describe the two-timescale environment.
  A bi-level off-policy RL framework is proposed accordingly, where two agents FTA and STA are set up for the FTCDs and STDDs respectively and are both trained with off-policy RL algorithms to exploit the samples efficiently.
  \item To cope with the non-stationary challenge of learning the two agents in two different timescales, our bi-level off-policy RL framework conduct coordination between STA and FTA instead of training them separately.
  In this context, we propose a technique called multi-timescale off-policy correction (MTOPC).
  With MTOPC, the bias of STA learning under the disturbance of FTA can be effectively eliminated.
  Such factors make the application of off-policy RL algorithms available for the STA.
  \item For the FTAs, the soft actor-critic (SAC) with continuous actions is adapted to learn a stochastic VVC policy; and for the STAs, we develop a multi-discrete soft actor-critic (MDSAC) algorithm to reduce the complexity of training and improve the efficiency.
  Comparing with the state-of-art RL algorithms, MDSAC can produce discrete action values for all STDDs simultaneously but alleviating the curse of dimensionality challenge.
\end{enumerate}

The rest of this paper is organized as follows.
\Cref{sec:prelim} formulates the two-timescale VVC problem in this paper, and also introduces key concepts and basic methods of RL as preliminaries.
Then in \cref{sec:methods}, the details of the proposed bi-level off-policy RL algorithm are derived and presented, and a two-timescale VVC is developed accordingly.
Moreover, in \cref{sec:numerical}, the results of the numerical study on the proposed two-timescale VVC are shown and analyzed.
Finally, \cref{sec:conclusion} concludes this paper.

\section{Preliminaries}
\label{sec:prelim}
In this section, we firstly formulate the two-timescale VVC problem in this paper.
Then, the settings of Markov decision process and its variants used in this paper is introduced.
In the last subsection, we cover the preliminaries of reinforcement learning and actor-critic framework to support \cref{sec:methods}.

\subsection{Two-timescale VVC Problem Formulation}
\label{sub:vvcform}
In this paper, we consider an ADN with ${n+1}$ nodes.
It can be depicted by an undirected graph $\mathcal{G}(\mathcal{N},\mathcal{E})$ with the collection of all nodes $\mathcal{N} = {0,...,n}$ and the collection of all branches $\mathcal{E} = {(i,j)\in \mathcal{N}\times\mathcal{N}}$.
The point of common coupling (PCC) is located at node 0 with a substation connected to the power grid simulated by a generation.

Both STDDs and FTCDs are installed in the ADN.
STDDs include $n_\text{OLTC}$ OLTCs and $n_\text{CB}$ CBs.
The tap of $i$th OLTC is $T_{O,i}$ and the tap of $i$th CB is $T_{B,i}$.
Typically, the number of taps are odd integers $\overline{T^i_{O}},\overline{T_{B,i}}$.
FTCDs include $n_\text{DG}$ DGs and $n_\text{SVC}$ SVCs. The reactive power of $i$th DG is $Q_{G,i}$ and that of $i$th SVC is $Q_{S,i}$.
Without loss of generality, we assume that all STDDs and FTCDs are installed on different nodes in $\mathcal{N}$.

In a slow-timescale VVC, the taps of OLTCs and CBs are optimized following the objective in \cref{eq:oslow} \cite{wangSafeOffpolicyDeep2019}.
$T$ is the number of slow-timescale VVC steps in one day and $t$ is one of the steps.
$P_\text{loss}$ is the active power loss of the ADN, and $T_\text{O, loss}, T_\text{B, loss}$ are the acting loss of OLTCs and CBs accordingly.
We have $T_\text{O, loss}^{(\tilde{t})} = \sum_{i=1}^{n_\text{OLTC}}|T^{(\tilde{t})}_{O, i} - T^{(\tilde{t}-1)}_{O, i}|$ and $T_\text{B, loss}^{(\tilde{t})} = \sum_{i=1}^{n_\text{CB}}|T^{(\tilde{t})}_{B, i} - T^{(t-1)}_{B, i}|$ when $t>0$ and $T_\text{O, loss}(0)=T_\text{B, loss}(0)=0$.
$C_P, C_O, C_B$ are the price coefficients of $P_\text{loss}, T_\text{O, loss}, T_\text{B, loss}$ accordingly.
$\underline{V}, \overline{V}$ are the voltage lower and upper limit, and $V_i$ is the voltage at node $i$.
\begin{equation}
  \begin{split}
    \label{eq:oslow}
    O_{S} &= \min\sum_{\tilde{t}=0}^{\tilde{T}-1} \left[ C_P P_\text{loss}^{(\tilde{t})} + C_O T_\text{O, loss}^{(\tilde{t})} + C_B T_\text{B, loss}^{(\tilde{t})} \right] \\
    & s.t. \;\;\;\; \underline{V} \leq V_i^{(\tilde{t})} \leq \overline{V}, \,\, \forall i \in \mathcal{N}, \tilde{t} \in [0, \tilde{T})
  \end{split}
\end{equation}

As for the fast-timescale VVC, $Q_G,Q_S$ are optimized given the tap settings from the slow timescale.
Let $T$ be the number of fast-timescale VVC steps in one day, and $t$ be one of the steps.
Then the problem is formulated as \cref{eq:ofast} \cite{xuAcceleratedADMMBasedFully2020}.
\begin{equation}
  \begin{split}
    \label{eq:ofast}
    O_{F} &= \min\sum_{t=0}^{T-1} \left[ C_P P_\text{loss}^{(t)} \right] \\
    & s.t. \;\;\;\; \underline{V} \leq V_i^{(t)} \leq \overline{V}, \,\, \forall i \in \mathcal{N},t \in [0, T)
  \end{split}
\end{equation}

The DGs are typically designed with redundant rated capacity for safety reasons and operate under maximum power point tracking (MPPT) mode.
Hence, the controllable range of the reactive power of DGs can be determined by the rated capacity $S_{G,i}$ and maximum power output $\overline{P_{G,i}}$.
The reactive power range of controllable devices is $|Q_{G,i}| \leq \sqrt{S_{G,i}^2-\overline{P_{G,i}}^2}$.
Also, we have $\underline{Q_{C,i}} \leq Q_{C,i} \leq \overline{Q_{C,i}}$ where $\underline{Q_{C,i}},\overline{Q_{C,i}}$ are the bounds of the $i$th SVC.

Because STDDs and FTCDs both exists in the ADN and need to be coordinated properly, the two problems \cref{eq:oslow,eq:ofast} are combined in this paper as \cref{eq:otwo} assuming that $T$ is a integer multiple of $\tilde{T}$ and $k=T / \tilde{T}$.
\begin{equation}
  \begin{split}
    \label{eq:otwo}
    O_{T} = \min_{T_{O},T_{B}}&\sum_{\tilde{t}=0}^{\tilde{T}-1} \Big[ C_O T_\text{O, loss}^{(k\tilde{t})} + C_B T_\text{B, loss}^{(k\tilde{t})}\\
    &+ C_P \min_{Q_G, Q_C}\sum_{\tau=0}^{k-1} P_\text{loss}^{(k\tilde{t}+\tau)}  \Big] \\
    & s.t. \;\;\;\; \underline{V} \leq V_i^{(k\tilde{t}+\tau)} \leq \overline{V},\\
    & \,\, \forall i \in \mathcal{N},  t \in [0, T), \tau \in [0, k)
  \end{split}
\end{equation}

Note that in a model-based optimization method, the VVC problems including \cref{eq:oslow,eq:ofast,eq:otwo} are solved with power flow constraints.
In this paper, we focus on a situation that the accurate power flow model is not available.

\subsection{Markov Decision Process and Reinforcement Learning}
\label{sub:mdp}
A fundamental assumption of reinforcement learning is that the environment can be described as an MDP.
The classic definition of an MDP is shown in \cref{def:mdp}.
\begin{definition}[Markov Decision Process]
  \label{def:mdp}
  A Markov decision process is a tuple $(\mathcal{S},\mathcal{A},p,r,\rho_0)$, where
  \begin{itemize}
    \item $\mathcal{S}$ is the state space,
    \item $\mathcal{A}$ is the action space,
    \item $p:\mathcal{S}\times\mathcal{A}\times\mathcal{S}\rightarrow \mathbb{R}^+=[0,\infty)$ is the transition probability distribution of the next state $s'\in\mathcal{S}$ at time $t+1$ with the current state $s\in\mathcal{S}$ and the action $a\in\mathcal{A}$ at time $t$,
    \item $r: \mathcal{S}\times\mathcal{A}\times\mathcal{S}\rightarrow \mathbb{R}$ is the immediate reward received after transiting from state $s$ to $s'$ due to action $a$,
    \item $\rho_0:\mathcal{S}\rightarrow \mathbb{R}^+$ is the probability distribution of the initial state $s_0$.
  \end{itemize}
\end{definition}

In the standard continuous control RL setting, an agent interacts with an environment (MDP) over periods of time according to a policy $\pi$.
The policy can be either deterministic, which means $a = \pi(s)$, or stochastic, which means $a \sim \pi(\cdot|s)$.
In this paper, stochastic policies are adapted.
From the definition of MDP, we can tell that if the environment is stationary, the cumulative reward can be improved by optimizing $\pi$.
Classically, the objective of RL at time $t$ is to maximize the expectation of the sum of discounted rewards $G_t = \sum_{i=t+1}^T \gamma^i r(s_i,a_i,s_{i+1})$
where $\gamma\in[0,1)$ is the discount factor, and $T$ is the length of episode.
A well-performing RL algorithm will learn a good policy $\pi$ from ideally minimal interactions with the environment,
as $\max J(\pi) = \mathop{\mathbb{E}}\limits_{\tau\sim\pi}\left[ G(\tau) \right]$.
Here $\tau$ is a trajectory of states and actions noted as $\{s_0,a_0,s_1,a_1,\dots,s_{T-1},a_{T-1},s_T\}$ and $G(\tau) = G_0, (s,a) \in \tau$.
$\tau \sim \pi$ is a trajectory with $\pi$ applied.

To get an optimal policy, one has to evaluate the policy $\pi$ under unknown environment transition dynamics $p$ and conduct improvement.
In reinforcement learning, such evaluation is carried out by defining two value functions $V^\pi(s)$ and $Q^\pi(s,a)$ as shown in \cref{eq:vq}.
$V^\pi(s)$ is the state-value function representing the expected discounted reward after state $s$ with the policy $\pi$.

$Q^\pi(s,a)$ is the state-action value function representing the expected discounted reward after taking action $a$ at state $s$ with the policy $\pi$.
\begin{equation}
    \label{eq:vq}
    \begin{split}
        V^{\pi}(s) &= \mathop{\mathbb{E}}\limits_{\tau \sim \pi}\left[\sum_{t=0}^T \gamma^t r_t \left| s_0 = s\right.\right] \\
        Q^{\pi}(s,a) &=  \mathop{\mathbb{E}}\limits_{\tau \sim \pi}\left[\sum_{t=0}^T \gamma^t r_t \left| s_0 = s, a_0 = a\right. \right] \\
        V^{\pi}(s) &= \mathop{\mathbb{E}}\limits_{a\sim \pi}Q^\pi(s,a)
    \end{split}
\end{equation}

According to the Bellman theorem and the Markov feature of MDP, the value functions can be recursively derived as \cref{eq:td},
\begin{equation}
    \label{eq:td}
    \begin{split}
    V^{\pi}(s) &= \mathop{\mathbb{E}}\limits_{\substack{a \sim \pi(\cdot|s) \\ s'\sim p(\cdot|s,a)}} \left[{r(s,a,s') + \gamma V^{\pi}(s')}\right] \\
    Q^{\pi}(s,a) &= \mathop{\mathbb{E}}\limits_{s'\sim p(\cdot|s,a)}
    \left[
        r(s,a,s') + \gamma \mathop{\mathbb{E}}\limits_{a'\sim \pi(\cdot|s')}{Q^{\pi}(s',a')}
    \right]
    \end{split}
\end{equation}
where $s'$ is the next state of $s$, and $a'$ is the next action. With \cref{eq:td}, the value functions can be relaxed from the whole trajectory but updated with batch of transitions.

\section{Methods}
\label{sec:methods}
In this section, we propose a novel bi-level off-policy reinforcement learning algorithm to solve the two-timescale VVC in ADNs. A summarized version of the RL-based two-timescale VVC is presented in \cref{sub:algo}.

We firstly propose a variant of MDP called bi-level Markov decision process (BMDP) in \cref{sub:bmdp} to describe the environment in two timescales.
The two-timescale VVC problem in \cref{sub:vvcform} is further formulated into BMDP accordingly.

Then, as shown in \cref{fig:proposed,fig:bmdp}, two agents FTA and STA are set up for the two timescales.
We adopt the well-known off policy RL algorithm SAC for FTA, which is described in \cref{sub:sac}.
To alleviate the curse of dimensionality challenge and improve the efficiency of STA, \cref{sub:mdsac} propose a novel algorithm MDSAC which allow the STA to decide actions for all STDDs simultaneously with outstanding sample efficiency.

Finally, instead of simply training FTA and STA separately, we calculate the exchanging factors between FTA and STA by our innovated MTOPC technique in \cref{sub:mtopc}.
It allows the FTA and STA trained together to optimize BMDP in a stationary manner based on the inherent Markov property of BMDP.

\subsection{Two-timescale VVC in Bi-level Markov Decision Process}
\label{sub:bmdp}
The standard RL setting in \cref{sub:mdp} considers only a single timescale, which does not match the two-timescale VVC problem in \cref{sub:vvcform}.
Hence, BMDP is defined as a variant of MDP in \cref{def:bmdp} to describe the environment (two-timescale VVC problem).
The transition probability $p$ is still unknown to the agents.
Note that though similar settings exist in some previous works related to hierarchical RL such as \cite{suttonMDPsSemiMDPsFramework1999}, BMDP is specially reformed for the two-timescale setting.
\begin{definition}[Bi-level Markov Decision Process]
  \label{def:bmdp}
  A bi-level Markov decision process is a joint of two MDPs defined in \cref{def:mdp} in two timescales, and is defined as a tuple $(k,\mathcal{S},\mathcal{A}_s,\mathcal{A}_f,r_s,r_l,p,\rho_0)$. Most of the symbols follow \cref{def:mdp}, and $\mathcal{A} = \mathcal{A}_s \cup \mathcal{A}_f$. The incremental parts are described as follows.
  \begin{itemize}
    \item $\mathcal{A}_s$ is the slow action space,
    \item $\mathcal{A}_f$ is the fast action space,
    \item $k$ is the timescale ratio and the $\mathcal{A}_s$ is only available every $k$ steps of the $\mathcal{A}_f$,
    \item when $t\mod{k}\neq 0$, the fast action $a\in\mathcal{A}_f$ makes that $s'\sim p(\cdot|s,a)$,
    \item when $t\mod{k}=0$, two transitions happens consequently: 1) the slow action $\tilde{a}\in\mathcal{A}_s$ is applied and the state transits from $\tilde{s}$ to $s$ as $s\sim p(\cdot|\tilde{s},\tilde{a})$; 2) the fast action $a\in\mathcal{A}_f$ is applied and $s'\sim p(\cdot|s,a)$, 
    \item $r_s: \mathcal{S}\times\mathcal{A}_s\times\mathcal{S}\rightarrow \mathbb{R}$ is the immediate reward received in slow timescale after transiting from state $\tilde{s}$ to $\tilde{s}'$ due to action $\tilde{a}$,
    \item $r_f: \mathcal{S}\times\mathcal{A}_f\times\mathcal{S}\rightarrow \mathbb{R}$ is the immediate reward received in fast timescale after transiting from state $s$ to $s'$ due to action $a$.
  \end{itemize}
\end{definition}
\begin{figure}[h]
  \centering
  \includegraphics[width=0.99\linewidth]{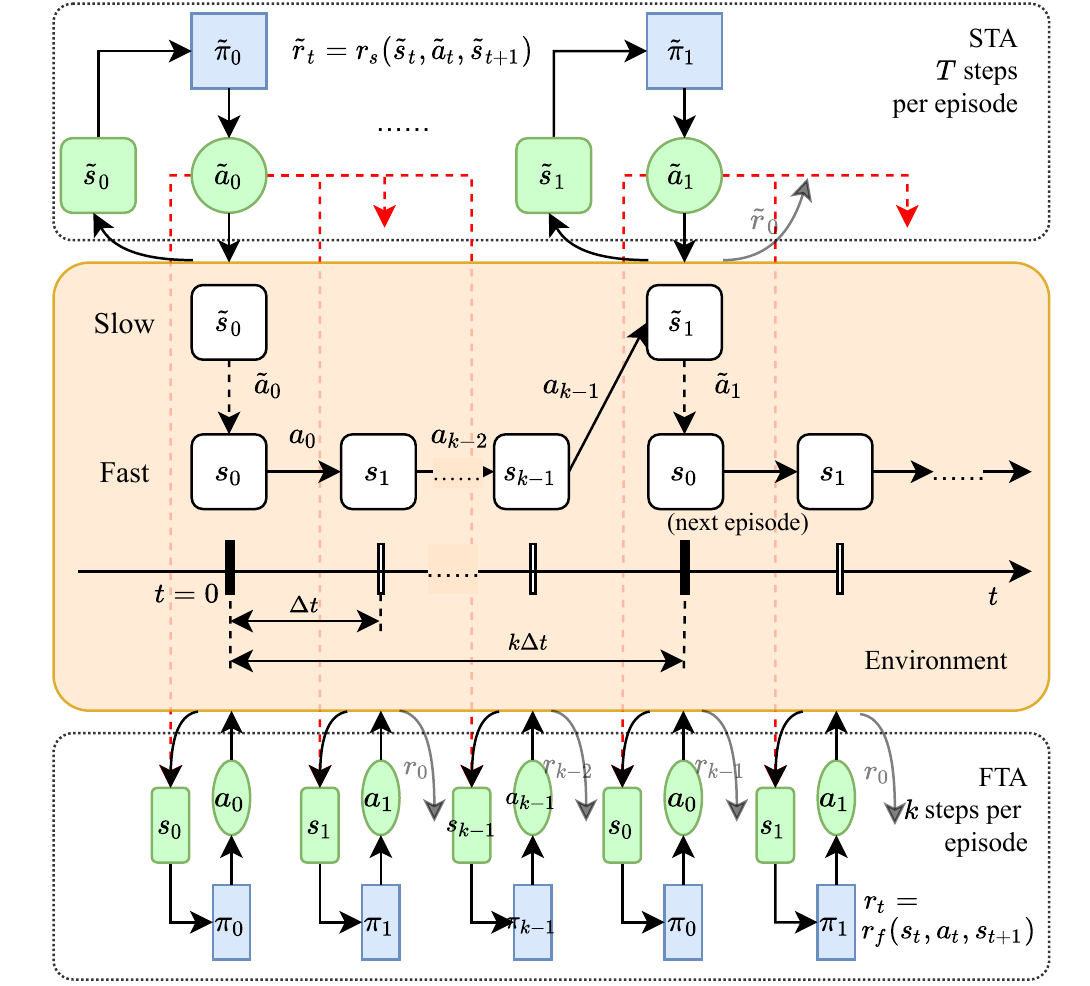}
  \caption{The setting of BMDP and the agents STA, FTA in this paper.}
  \label{fig:bmdp}
\end{figure}

Orange painted part in \cref{fig:bmdp} illustrates BMDP step by step.
The setting of BMDP has ensured the environment is Markovian from the perspective of bi-level.
In each episode ($k$ steps) of the fast layer, the initial state depends on the certain slow action.
We therefore has to include the corresponding slow action in the state explicitly, which is emphasized by red dashed lines.
The slow actions $\tilde{a}_0, \tilde{a}_1, \cdots$ can be seen as the exchanging factors from STA to FTA as shown in \cref{fig:proposed}.

Obviously, the transition from $\tilde{s}_0$ to $\tilde{s}_1$ depends not only on the slow action $\tilde{a}_0$, but also on the $k$ fast actions $a_0, a_1, \cdots, a_{k-1}$.
Hence, in the view of slow level, the environment does not satisfy Markov property.
In another way, \textbf{BMDP is Markovian bi-levelly in fast level, but non-Markovian in slow level}.
In the following \cref{sec:methods}, one of the major ideas is to take full advantage of the inherent Markov property of BMDP and carry out stable learning and control process in both timescales.

To formulate the two-timescale VVC problem \cref{eq:otwo} into BMDP,
the specific definitions of episodes, state spaces, action spaces and reward functions are designed as follows.

\subsubsection{Episode}
An episode of STA is defined as one day, and the step size $\Delta t$ is one hour. We have $\tilde{T}=24$ total steps in one episode. An episode of FTA includes $k=12$ steps in one STA step with $\Delta \hat t = 5 \text{min}$. Note that $T,k,\Delta t,\Delta \hat t$ are all alterable while satisfying $\Delta t = k\Delta \hat t$.

\subsubsection{State Space}
The common state of BMDP is defined as a vector $(\mathbf{P},\mathbf{Q},\mathbf{V},\textbf{T}_O,\textbf{T}_B,t)$, where
$\mathbf{P},\mathbf{Q}$ are the vectors of nodal active/reactive power injections $P_i, Q_i(\forall i \in \mathcal{N})$,
$\mathbf{V}$ is the vector of voltage magnitudes $V_i(\forall i \in\mathcal{N})$,
$\mathbf{T}_O=[T_{O,1},\cdots,T_{O,n_\text{OLTC}}]^T$ is the vector of OLTC taps,
$\mathbf{T}_B=[T_{B,1},\cdots,T_{B,n_\text{CB}}]^T$ is the vector of CB taps,
$t$ is the time in one day.

\subsubsection{Action Spaces}
For FTA, the action space $\mathcal{A}_f$ includes all the controllable reactive power of FTCDs, that is, $[Q_{G,1}, \cdots, Q_{G,n_\text{DG}}, Q_{C,1}, \cdots, Q_{C,n_\text{SVC}}]^T$.
Since the reactive power generations are continuous in \cref{sub:vvcform}, $\mathcal{A}_f$ is defined as a box space with the lower bound $[-\overline{Q_{G,1}}, \cdots, -\overline{Q_{G,n_\text{DG}}}, \underline{Q_{C,1}}, \cdots, \underline{Q_{C,n_\text{SVC}}}]$, and upper bound $[\overline{Q_{G,1}}, \cdots, \overline{Q_{G,n_\text{DG}}}, \overline{Q_{C,1}}, \cdots, \overline{Q_{C,n_\text{SVC}}}]$, where $\overline{Q_{G,i}}=\sqrt{S_{G,i}^2-\overline{P_{G,i}}^2}$.

For STA, the action space $\mathcal{A}_s$ includes all the tap settings of STDDs, that is, $(\mathbf{T}_O,\mathbf{T}_B)$.
Each tap setting $T_{O,i}$ or $T_{B,i}$ can be selected from $\overline{T_{O,i}}$ or $\overline{T_{B,i}}$ taps, so $\mathcal{A}_s$ is a discrete space with multi dimensions, also known as multi-discrete space.
The dimensions are listed in the vector $[\overline{T_{O,1}},\cdots,\overline{T_{O,n_\text{OLTC}}},\overline{T_{B,1}},\cdots,\overline{T_{B,n_\text{CB}}}]^T$.

\subsubsection{Reward Functions}
The reward functions maps a certain transition to a single value to cumulate and maximize.

For the fast timescale, note a transition as $(s, a, s')$ where $s, s'\in\mathcal{S}$ and $a\in\mathcal{A}_f$.
Then $r_f$ is defined as \cref{eq:rf} according to the inner minimization of \cref{eq:otwo}.
Here $[\cdot]_+$ is the rectified linear unit function defined as $[x]_+=\max(0, x)$, $C_V$ is a penalty multiplier for the voltage constraints, and $V_\text{loss}$ is a smooth index of the voltage violations called voltage violation rate (VVR).
\begin{align}
  \label{eq:rf}
  &r_f(s,a,s') = -C_P P_\text{loss}(s') - C_V V_\text{loss}(s')\\
  \label{eq:ploss}
  &P_\text{loss}(s') = \sum_{i\in\mathcal{N}}P_i(s'), \\
  \label{eq:vloss}
  &V_\text{loss}(s') = \sqrt{\sum_{i\in\mathcal{N}}\left[[V_i(s')-\overline{V}]_{+}^2 + [\underline{V}-V_i(s')]_{+}^2\right]}
\end{align}

For the slow timescale, the transition is noted as $(\tilde{s}, \tilde{a}, \tilde{s}')$,
where $\tilde{s}, \tilde{s}' \in \mathcal{S}$ and $\tilde{a} \in \mathcal{A}_s$.
Also, as shown in \cref{fig:bmdp}, we have as series of fast timescale samples between $\tilde{s}$ and $\tilde{s}'$. Note them as $s_0,a_0,s_1,a_1,\cdots,s_{k-1},a_{k-1},s_k$ where $s_k=\tilde{s}'$.
Since the objective of the STA considers switching cost of STDDs, the reward function of the slow timescale $r_s$ includes the switching cost part and the cumulative reward of the fast timescale.
\begin{equation}
  \begin{aligned}
    \label{eq:rs}
    r_s(\tilde{s},\tilde{a},\tilde{s}',\{s_\tau,&a_\tau|\tau\in[0,k)\},s_k) = \\
    &- C_O T_\text{O,loss}(\tilde{s},\tilde{s}') -C_B T_\text{B,loss}(\tilde{s},\tilde{s}')\\
    &- R_f(\{s_\tau,a_\tau|\tau\in[0,k)\},s_k)
  \end{aligned}
\end{equation}
\vspace{-8pt}
\begin{align}
  \label{eq:toloss}
  &T_\text{O,loss}(\tilde{s},\tilde{s}') = \sum_{i=1}^{n_\text{OLTC}}|T_{O, i}(\tilde{s}') - T_{O, i}(\tilde{s})|\\
  \label{eq:tbloss}
  &T_\text{B,loss}(\tilde{s},\tilde{s}') = \sum_{i=1}^{n_\text{CB}}|T_{B, i}(\tilde{s}') - T_{B, i}(\tilde{s})|\\
  \label{eq:Rf}
  &R_f(\{s_\tau,a_\tau|\tau\in[0,k)\},s_k) = \sum_{\tau=0}^{k-1}r_f(s_\tau, a_\tau, s_{\tau+1})
\end{align}

To solve BMDP, two agents STA and FTA are set up for the slow and fast timescales respectively, as shown in \cref{fig:bmdp}.
The policy of STA is marked as ${\tilde{\pi}}$, while that of FTA is $\pi$.
An intuitive method would be training STA and FTA separately.
However due to the fact that the environment is non-Markovian from the view of STA, it violates the basic assumption of RL algorithms and leads to non-stationary learning process.
In \cref{sub:mtopc}, MTOPC is proposed addressing this challenge.

\subsection{Soft Actor-Critic for FTA}
\label{sub:sac}
SAC \cite{haarnojaSoftActorcriticAlgorithms2018} is a state-of-art off-policy RL method for MDPs with continuous action space.
It is implemented in the actor-critic framework, where the actor is the stochastic policy $\pi(\cdot|s)$, and the critic is the state-action value function $Q^\pi(s,a)$.
$\pi$ and $Q^\pi$ are both approximated by deep neural networks (DNN) with parameters noted as $\phi_f$ and $\theta_f$ in practice.

In SAC, the definition of state-action value function is entropy-regularized as
\begin{equation}
    Q^{\pi}_f(s,a) = \mathop{\mathbb{E}}\limits_{\tau \sim \pi} \bigg[\sum_{t=0}^T \gamma^t r_t + \alpha_f\sum_{t=1}^T \gamma^t H\big(\pi(\cdot|s_t)\big) \big| s_0 = s, a_0 = a \bigg]
\end{equation}
where $H(\pi(\cdot|s_t))=\mathop{\mathbb{E}}\limits_{a\sim\pi(\cdot|s_t)}\left[-\log \pi(\cdot|s_t)\right]$ is the entropy for the stochastic policy at $s_t$, $\alpha_f$ is the temperature parameter.

During the learning process, all the samples of MDP are stored in the replay buffer $\mathcal{D}$ as $(s,a,r_f,s')\in \mathcal{D}$.
To approximate $Q^{\pi}_f$ iteratively, Bellman equation is applied to the entropy-regularized $Q^{\pi}_f$ as \cref{eq:bellman} since $\mathbb{E}_{s', a'}H\left(\pi\left(\cdot|s'\right)\right)=-\mathbb{E}_{s', a'}\log \pi\left(\cdot|s'\right)$. 
\begin{equation}
  \label{eq:bellman}
  \begin{aligned}
    &Q^{\pi}_f(s,a) \approx
    \mathop{\mathbb{E}}\limits_{\substack{s'\sim p(\cdot|s,a)\\ a'\sim \pi(\cdot|s')}}
    \Big[r_f + \gamma \left(Q^{\pi}_f\left(s',a'\right) + \alpha_f H\left(\pi\left(\cdot|s'\right)\right)\right) \Big] \\
    & = \mathop{\mathbb{E}}\limits_{\substack{s'\sim p(\cdot|s,a)\\ a'\sim \pi(\cdot|s')}} \Big[r_f + \gamma \left( Q^{\pi}_f\left(s', a'\right) - \alpha_f \log \pi \left(\left. a'\right|s' \right)  \right)\Big]
  \end{aligned}
\end{equation}
Then, mean-squared Bellman error (MSBE) \cref{eq:qloss} is minimized to update the $Q$ network. Note $y=r_f + \gamma \Big( Q^{\pi}_f\left(s', a'\right) - \alpha_f \log \pi \left(\left. a'\right|s' \right)  \Big)$. All expectations are approximated with Monte Carlo method.
\begin{equation}
  \label{eq:qloss}
  L_f^Q(\phi_f) = \mathop{\mathbb{E}}\limits_{(s,a,r_f,s')\in\mathcal{D}} [Q_{\phi_f}-y]^2
\end{equation}

The policy is optimized to maximize the state value function $V^\pi(s)=\mathop{\mathbb{E}}\limits_{a\sim \pi}Q^\pi(s,a)$. In SAC, the reparameterization trick using a squashed Gaussian policy is introduced: $a'_{\theta_f}(s, \xi) = \tanh\left( \mu_{\theta_f}(s) + \sigma_{\theta_f}(s) \odot \xi \right), \, \xi \sim \mathcal{N}(0, \mathbf{I})$, where $\mu_{\theta_f},\sigma_{\theta_f}$ are two DNNs. Hence, and the policy can be optimized by minimizing \cref{eq:piloss}, where $a' = a'_{\theta_f}(s,\xi)$.
\begin{equation}
  \label{eq:piloss}
  L^\pi_f(\theta_f) = - \mathop{\mathbb{E}}\limits_{\substack{s\sim \mathcal{D}\\ \xi\sim \mathcal{N}}} \left[ Q^\pi\left(s,a'\right) - \alpha_f \log \pi_{\theta_f}\left(\left.a'\right|s\right) \right]
\end{equation}

Other practical techniques in \cite{haarnojaSoftActorcriticAlgorithms2018} are omitted here due to limited space.

\subsection{Multi-Discrete Soft Actor-Critic for STA}
\label{sub:mdsac}
Comparing with the FTA with continuous actions, STA has discrete actions in multi dimensions.
It makes fundamental differences in the RL algorithm.
Therefore, we propose a variant of SAC called multi-discrete soft actor-critic (MDSAC) as follows.
MDSAC fully suits the multi-discrete action space of STA and can carry out high efficiency training in an off-policy manner.

\subsubsection{Policy DNN Architecture}
Because the policy should select an action for each of the STDDs, we design a multi-head DNN for ${\tilde{\pi}}$.
Note $n_s = n_\text{CB} + n_\text{OLTC}$ as the number of STDDs, and $m_i$ as the number of taps of $i$th STDD.
As shown in \cref{fig:policy}, $\tilde{s}$ is mapped to a shared representation, which are passed to $n_s$ heads with hidden layers.
The $i$th head produces $m_i$ digits, passes them to the softmax layer, and receives a vector of $m_i$ digits ${\tilde{\pi}}^i(\tilde{s}) = [{\tilde{\pi}}^i(\tilde{a}^i_1, \tilde{s}),\cdots,{\tilde{\pi}}^i(\tilde{a}^i_{m_i}, \tilde{s})]^T$, where $\tilde{a}^i_1,\cdots,\tilde{a}^i_{m_i}$ are all the possible actions of $i$th STDD.
Note that $\sum_{j=1}^{m_i}{\tilde{\pi}}^i_j(\tilde{s})=1$.
Finally, the probability value of a certain action $\tilde{a}$ is ${\tilde{\pi}}(\tilde{a}|\tilde{s}) = \prod_{i=1}^{n_s}{\tilde{\pi}}^i(\tilde{a}^i, \tilde{s})$.

\subsubsection{Value DNN Architecture}
One of the most challenging problem in off-policy RL with multi-discrete action space is that the state-action value function is non-trivial to implement.
Classically, if the action space is discrete, the state-action value function is designed as $Q:\mathcal{S}\rightarrow \mathbb{R}^m$, where $m$ is the number of actions.
However, with multiple devices, it upgrades to $Q_s:\mathcal{S}\rightarrow \mathbb{R}^{m_1}\times\mathbb{R}^{m_2}\cdots\times\mathbb{R}^{m_{n_s}}$ mathematically.
It means that $Q_s$ has $\prod_{i=1}^{n_s}m_i$ outputs.
When $n_s$ increases, the cost of memory and CPU time will grow exponentially.
Worse still is that bloated $Q_s$ outputs requires much more samples to train, which is unaffordable in ADNs.
Such exponential complexity can be seen as the curse of dimensionality problem and hinders the implementation of STA in practice.

To alleviate the problem above, we introduce a device decomposition technique to the value DNN architecture inspired by \cite{sunehagValueDecompositionNetworksCooperative2017}, where the value function is relaxed as the sum of independent value functions with local states and actions of each agent in multi-agent settings.
A limitation of \cite{sunehagValueDecompositionNetworksCooperative2017} is that such sum combination may reduce the approximation performance of the DNN.
In MDSAC, we introduce $\tilde{s}$-adaptive affine parameters $c(\tilde{s})$ to address the limitation.

\begin{figure}[h]
  \centering
  \subfloat[The architecture of policy neural network ${\tilde{\pi}}(\cdot|\tilde{s})$. \label{fig:policy}]{%
  \includegraphics[width=1\linewidth]{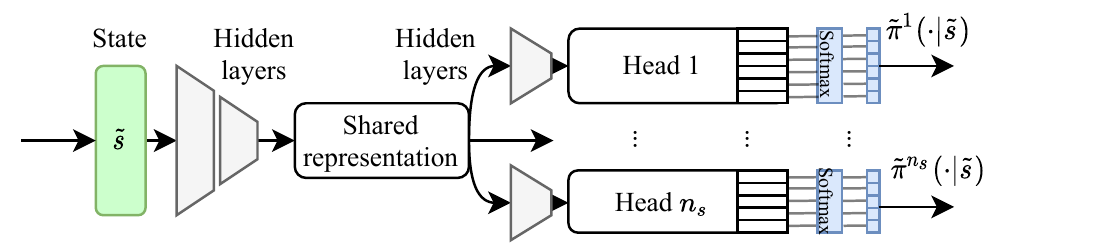}}
  \vfill
  \subfloat[The architecture of state-action value neural network $Q_s$. \label{fig:value}]{%
  \includegraphics[width=1\linewidth]{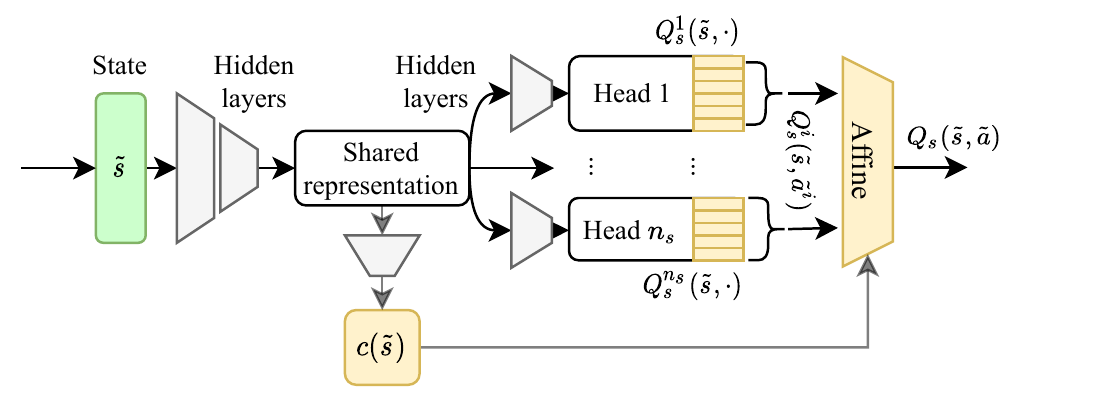}}
  \caption{DNN architecture design of ${\tilde{\pi}}$ and $Q_s$.}
  \label{fig:mdsac}
\end{figure}

As shown in \cref{fig:value}, the share representation is fed to two parts: 1) the $i$th head produces a vector $Q_s^i(\tilde{s})$ of $m_i$ scalars noted as $Q_s^i(\tilde{s},\tilde{a}^i_j), j\in[1,m_i]$; 2) a vector $c(\tilde{s})$ containing $n_s+1$ scalars, where $c(\tilde{s}) = [c_0,c_1,\cdots,c_{n_s}]^T$.
For a certain action $\tilde{a}$, the device-wise state-action values are selected as $Q_s^1(\tilde{s}, \tilde{a}^1),\cdots,Q_s^{n_s}(\tilde{s}, \tilde{a}^{n_s})$.
Then, they are combined in an affine mixing network with ratios $c(\tilde{s})$, as \cref{eq:qcomb}.
Because the ratios $c(\tilde{s})$ are also generate from DNNs, the flexibility of approximation to actual $Q_s$ is generally boosted comparing with \cite{sunehagValueDecompositionNetworksCooperative2017}.
Also, $c_0(\tilde{s})$ learns a base value for each state, which is inherently similar to the well-known dueling network architecture \cite{wangDuelingNetworkArchitectures2016}.
\begin{equation}
  \label{eq:qcomb}
  Q_s(\tilde{s},\tilde{a}) = c_0(\tilde{s}) + \sum_{i=1}^{n_s} c_i(\tilde{s}) Q_s^i(\tilde{s},\tilde{a}^i)
\end{equation}

\subsubsection{Updates of Actor and Critic}
Since the action space is discrete, the policy maps $\tilde{s}$ to the probability value of action $\tilde{a}$ instead of a probability density function.
As a result, the expectation on the policy ($\mathop{\mathbb{E}}_{\tilde{a}'\sim \tilde{\pi}(\cdot|\tilde{s}')}$) can be calculated explicitly instead of in a Monte Carlo way.
That means the state value function can be expressed as
\begin{align}
  V^{\tilde{\pi}}_s(\tilde{s}) &= \mathop{\mathbb{E}}\limits_{\tilde{a}\sim {\tilde{\pi}}(\cdot|\tilde{s})} \left[Q_s(\tilde{s},\tilde{a}) - \alpha_s \log {\tilde{\pi}}(\tilde{a}|\tilde{s})\right] \\
  &= \sum_{i=1}^{n_s} {\tilde{\pi}}^i(\tilde{s})^T \left[ c_i(\tilde{s}) Q^i_s (\tilde{s}) - \alpha_s \log {\tilde{\pi}}^i(\tilde{s})\right] + c_0(\tilde{s})
\end{align}
where the linearity of expectation operator and the mixing network are leveraged.

From \cref{eq:vq,eq:td}, we have
\begin{equation}
  \label{eq:qs}
  Q_s^{\tilde{\pi}}(\tilde{s},\tilde{a}) = \mathop{\mathbb{E}}\limits_{\tilde{s}'\sim \tilde p(\cdot|\tilde{s},\tilde{a})}[r_s(\tilde{s},\tilde{a},\tilde{s}') + \gamma V_s^{\tilde{\pi}}(\tilde{s}')]
\end{equation}
where $\tilde p(\cdot|\tilde{s},\tilde{a})$ is the current environment transition probability distribution.
Note ${\tilde{y}}(\tilde{s}'|\tilde{s},\tilde{a})=r_s(\tilde{s},\tilde{a},\tilde{s}') + \gamma V_s^{\tilde{\pi}}(\tilde{s}')$.
If $\tilde p(\cdot|\tilde{s},\tilde{a})$ holds, the expectation can be calculated by Monte Carlo method on $\mathcal{D}$.
However, $\tilde p(\cdot|\tilde{s},\tilde{a})$ is changing with $\pi$ of FTA varying in \cref{sub:sac}.
This is ignored temporarily here and will be corrected in \cref{sub:mtopc}.
\begin{equation}
  \label{eq:mdq}
  L_s^Q(\phi_s) = \mathop{\mathbb{E}}\limits_{(\tilde{s},\tilde{a},\tilde{s}',r_s)\sim \mathcal{D}} \left[ {\tilde{y}}(\tilde{s}'|\tilde{s},\tilde{a}) - Q_{\phi_s}(\tilde{s},\tilde{a}) \right]^2
\end{equation}
\begin{equation}
  \label{eq:mdpi}
  L_s^\pi(\theta_s) = -\mathop{\mathbb{E}}\limits_{S\sim \mathcal{D}} \left[
    V_s^{\tilde{\pi}}(\tilde{s})
  \right]
\end{equation}

Accordingly, the parameters $\phi_s$ of $Q_s$ and $\theta_s$ of ${\tilde{\pi}}$ can be optimized by mimimizing \cref{eq:mdpi,eq:mdq}.

\subsection{Multi-timescale Off-policy Correction}
\label{sub:mtopc}
As an off-policy RL algorithm, MDSAC stores all transitions in the experience replay buffer $\mathcal{D}$.
In \cref{eq:qs}, if $\tilde p(\cdots|\tilde{s},\tilde{a})$ holds as assumed by MDP, the expectation can be calculated with Monte Carlo method by sampling $\mathcal{D}$, like SAC in \cref{sub:sac} does.
However, in BMDP only $p(\cdot|s,a)$ is assumed to be stationary.
Mathematically, as shown in \cref{fig:bmdp}, the current probability of transition from $\tilde{s}_0$ to $\tilde{s}_1$ is
\begin{equation}
  \label{eq:currentp}
  \tilde p (\tilde{s}_1| \tilde{s}_0, \tilde{a}_0) = p(s_0|\tilde{s}_0,\tilde{a}_0)\prod_{i=0}^{k-1}\pi_i(a_i|s_i)p(s_{i+1}|s_i,a_i)
\end{equation}
where $s_k = \tilde{s}_1$.
While the FTA is training with \cref{eq:piloss}, $\pi$ is varying from time to time.
Hence, we have marked $\pi_0,\cdots,\pi_{k-1}$ for each FTA step.
Though STA needs to calculate the expectation on \cref{eq:currentp}, the data in $\mathcal{D}$ is sampled from another probability distribution,
\begin{equation}
  \label{eq:pastp}
  \tilde p^0 (\tilde{s}_1| \tilde{s}_0, \tilde{a}_0) = p(s_0|\tilde{s}_0,\tilde{a}_0)\prod_{i=0}^{k-1}\pi_i^0(a_i|s_i)p(s_{i+1}|s_i,a_i)
\end{equation}
where $\pi^0$ is the behavior policy used in past, and $\pi^0$ is different from the current policy $\pi$.
Obviously, the samples from $\mathcal{D}$ are not valid for direct Monte Carlo any more.
In another way, during the learning process, the past experience of STA is no longer correct for the current learning and can lead to significant bias.

To reuse the samples in $\mathcal{D}$ and leverage the off-policy MDSAC's high sample efficiency, we propose multi-timescale off-policy correction (MTOPC) based on importance sampling (IS) method.
IS is widely used in policy gradient RL algorithms like A2C \cite{mnihAsynchronousMethodsDeep2016} and PPO \cite{schulmanProximalPolicyOptimization2017}.

To estimate the expectation in \cref{eq:qs}, MTOPC is derived as \cref{eq:mtopc},
\begin{equation}
  \label{eq:mtopc}
  \begin{aligned}
  &\mathop{\mathbb{E}}\limits_{\tilde p(\cdot|\tilde{s},\tilde{a})}{\tilde{y}}(\tilde{s}'|\tilde{s},\tilde{a}) = \mathop{\mathbb{E}}\limits_{\tilde{p}^0(\cdot|\tilde{s},\tilde{a})} \frac{\tilde p(\tilde{s}'|\tilde{s},\tilde{a})}{\tilde{p}^0(\tilde{s}'|\tilde{s},\tilde{a})} {\tilde{y}}(\tilde{s}'|\tilde{s},\tilde{a})) \\
  &=\mathop{\mathbb{E}}\limits_{\tilde{p}^0(\cdot|\tilde{s},\tilde{a})} \frac{
    p(s_0|\tilde{s},\tilde{a})\prod_{i=0}^{k-1}\left[
      \pi(a_{i}|s_i) p(s_{i+1}|s_i,a_i)
  \right]}{
    p(s_0|\tilde{s},\tilde{a})\prod_{i=0}^{k-1}\left[
    \pi^0_i(a_{i}|s_i) p(s_{i+1}|s_i,a_i)
  \right]} {\tilde{y}}(\tilde{s}'|\tilde{s},\tilde{a}) \\
  &= \mathop{\mathbb{E}}\limits_{\tilde{p}^0(\cdot|\tilde{s},\tilde{a})} \prod_{i=0}^{k-1} \frac{\pi(a_{i}|s_i)}{\pi^0_i(a_{i}|s_i)}  {\tilde{y}}(\tilde{s}'|\tilde{s},\tilde{a}) \\
  &= \mathop{\mathbb{E}}\limits_{\tilde{p}^0(\cdot|\tilde{s},\tilde{a})} \omega \cdot {\tilde{y}}(\tilde{s}'|\tilde{s},\tilde{a}) = \mathop{\mathbb{E}}\limits_{\mathcal{D}} \omega \cdot {\tilde{y}}(\tilde{s}'|\tilde{s},\tilde{a})
\end{aligned}
\end{equation}
where $\pi$ is the current FTA policy, $\pi^0$ is the behavior (original) policies, and $\omega$ is the correction factor calculated by FTA,
\begin{equation}
  \label{eq:omega}
  \omega = \prod_{i=0}^{k-1} \frac{\pi(a_{i}|s_i)}{\pi^0_i(a_{i}|s_i)}.
\end{equation}
Simlar technique was also studied by \cite{nachumDataEfficientHierarchicalReinforcement2018} in hierarchical RL instead of the multi-timescale settings.

In practice, $\pi_0(a|s)$ is stored with the transition in $\mathcal{D}$, and $\pi(a|s)$ is calculated using the latest FTA policy. Because the cumulative product in \cref{eq:omega} may lead to high variance and numerical problem, we clip $\omega$ as $\omega'=\max(\min(\omega,\overline\omega), \underline\omega)$, where $\underline\omega),\overline\omega$ are the lower and upper bound for $\omega'$.

Accordingly, \cref{eq:mdq} is corrected as
\begin{equation}
  \label{eq:finalmdq}
  L_s^Q(\phi_s) = \mathop{\mathbb{E}}\limits_{\mathcal{D}} \left[ \omega'\cdot {\tilde{y}}(\tilde{s}'|\tilde{s},\tilde{a}) - Q_{\phi_s}(\tilde{s},\tilde{a}) \right]^2
\end{equation}
with batches of $(\tilde{s},\tilde{a},s_0,a_0,\cdots,s_{k-1},a_{k-1},\tilde{s}',r_s)$ sampled from $\mathcal{D}$ and the latest $\pi$.

\subsection{Two-timescale VVC with Bi-level RL}
\label{sub:algo}
The overall algorithm combining \cref{sub:bmdp,sub:sac,sub:mdsac,sub:mtopc} are summarized in \cref{alg:proposed}.
Note that the gradient steps of STA or FTA are carried out in parallel with the controlling process.
Typically, one gradient step is executed every one or several control steps.

\begin{algorithm}
  \caption{Two-timescale VVC with Bi-level RL}
  \label{alg:proposed}
  \SetKwProg{Parallel}{foreach}{ do in parallel}{end}
  Given learning rates $\sigma_s,\sigma_f$, temperature parameters $\alpha_s,\alpha_f$\;
  Initialize STA and FTA's policy and value functions' parameter vectors $\phi_s,\theta_s,\phi_f,\theta_f$\;
  $\mathcal{D}\leftarrow \emptyset$\;
  \ForEach{
    STA episode
  }{
    Get the initial state $\tilde{s}_0$, $t\leftarrow 0$\;
    \Parallel{
      STA step
    }{
      $\tilde{a}_t\sim{\tilde{\pi}}(\cdot|\tilde{s}_t)$\;
      Feed $\tilde{a}_t$ to the environment, get next state $s_0$\;
      $\mathcal{D}_t \leftarrow \emptyset$, $i\leftarrow 0$\;
      \ForEach{
        FTA step in a $k$-step episode
      }{
        $a_i\sim\pi(\cdot|s_i)$, $p_i\leftarrow\pi(a_i|s_i)$\;
        Feed $a_i$ to the environment, get reward $r_{f,i}$, next state $s_{i+1}$\;
        $\mathcal{D}_t\leftarrow\mathcal{D}_t\cup\{(s_i,a_i,r_{f,i},s_{i+1},p_i)\}$\;
        $i\leftarrow i+1$\;
      }
      Get reward $r_{s,t}$, state $\tilde{s}'\leftarrow s_{k}$\;
      $\mathcal{D}\leftarrow\mathcal{D}\cup\{(\tilde{s}_t,\tilde{a}_t,r_{s,t},\tilde{s}_{t+1}, \mathcal{D}_t)\}$\;
      $t\leftarrow t+1$\;
    }
    \Parallel{FTA gradient step}{
      Sample a batch of $(s,a,r_f,s')$ from $\mathcal{D}$\;
      Update $Q_{\phi_f}$ with \cref{eq:qloss}:\\
      \hfill $\phi_f\leftarrow \phi_f - \sigma_f\nabla_{\phi_f}L_f^Q(\phi_f)$\;
      Update $\pi_{\theta_f}$ with \cref{eq:piloss}:\\
      \hfill $\theta_f \leftarrow \theta_f - \sigma_f\nabla_{\theta_f} L_f^\pi(\theta_f)$\;
    }
    \Parallel{STA gradient step}{
      Sample a batch of $(\tilde{s},\tilde{a},r_s,\tilde{s}',D)$ from $\mathcal{D}$\;
      Calculate $\omega$ with \cref{eq:omega}\;
      Update $Q_{\phi_s}$ with \cref{eq:finalmdq}:\\
      \hfill $\phi_s\leftarrow \phi_s - \sigma_s\nabla_{\phi_s}L_s^Q(\phi_s)$\;
      Update $\pi_{\theta_s}$ with \cref{eq:mdpi}:\\
      \hfill $\theta_s \leftarrow \theta_s - \sigma_s\nabla_{\theta_s} L_s^\pi(\theta_s)$\;
    }
  }
\end{algorithm}

In real world application, the FTA and STA can be pretrained with an approximated model directly or with \cite{liuTwostageDeepReinforcement2020}.

\section{Numerical Study}
\label{sec:numerical}
With distribution test feeders, numerical studies are conducted to validate the advantage of the proposed two-timescale VVC based on bi-level off-policy RL.
The two-timescale VVC problem formulated in BMDP is implemented under the scheme of Gym toolkit \cite{brockmanOpenAIGym2016}.
A modified version of 33-bus distribution test feeder \cite{baranNetworkReconfigurationDistribution1989} is simulated as the target ADN.
In the 33-bus system, 4 DGs are connected to node 18,22,25,33 with 0.85 MVA rating, 1 OLTC is installed on the branch from node 1 to node 2, and 1 CB is installed on node 8.
The OLTC have 11 taps, and the ratio ranges evenly from 0.9 to 1.1.
Also, the CB have 11 taps and the output ranges from -1 MVar to 1 MVar evenly.
The profile of DGs and loads from a pilot project in eastern China is scaled to $[0,2]$ and allocated to the existing nodes in the standard case by multiplication.
All DGs are assumed to be installed with smart inverters following IEEE 1547.
The voltage limitations are set to be $[0.95, 1.05]$.
If the power flow does not converge or there is any $V_i > 1.15$ or $V_i < 0.85$ at certain time step, there is thought to be a grid failure and the episode is terminated with a negative reward $r_f = -500$.
As suggested in \cite{wangSafeOffpolicyDeep2019}, the electricity price $C_P$ is assumed to be \$40/MWh,
and the action price $C_O$ and $C_B$ are assumed to be \$0.1/tap.
The price of voltage violation rate is set to be \$100/$\text{p.u.}\cdot 5 \text{mins}$.

\subsection{Proposed and Baseline Algorithms Setup}
To compare with traditional two-timescale VVC, we implement an optimization-based benchmark as single-step mixed-integer conic programming (MICP) for STA and second order conic programming (SOCP) for FTA, which is equivalent to the method in \cite{jhaBiLevelVoltVAROptimization2019}.
Note that because the actual model is unknown to the operator, optimization methods are carried out with an approximated model.
In this paper, this benchmark is marked as ``A-OPT''.
The one with oracle model is marked as ``OPT''. But since the optimization is single step-wise, global optimality is not guaranteed.

Since works are still limited regarding DRL-based two-timescale VVC, we implement the method in \cite{yangTwoTimescaleVoltageControl2020} as the second benchmark called ``S-DQN''.
It adopts the classic DQN algorithm for STA by enumerating all possible STDDs combinations in the $Q$ network and has $\prod_{i=1}^{n_s}m_i$ actions.
For FTA, it adopts the SOCP method based on detailed ADN models.
Note that when we adopts the approximated model for SOCP, the STA cannot converge to a reasonable range in limited time (100k steps).
Hence, we implement S-DQN with oracle models, which is not required by our proposed method.

To show the effectiveness of the proposed MTOPC method, we add the third benchmark called ``Non-OPC''.
This benchmark is the same as the proposed one except for $\omega = 1$ instead of \cref{eq:omega}.

In the following numerical studies, the proposed method in \cref{alg:proposed} is marked as ``Proposed''.
The discrete variables $\mathbf{T}_O,\mathbf{T}_B$ in the state are encoded with one-hot embedding. That is, a certain tap like $T_{o,i}$ is transformed to a $\overline{T_{0,i}}$-length binary vector $[0,\dots,1,\dots,0]^T$ with zeros except for $1$ at $T_{o,i}$-th position.
The bound of MTOPC $\omega$ is set as $\underline{\omega}=0.1$, $\overline{\omega}=10$.
All DRL-based methods implemented are trained under PyTorch with the parameters of DNNs randomly initialized and updated in batches.
The batch size is chosen as 128 and the replay buffer is designed to be a queue with size of $2000\times 12=24000$ FTA steps.
All the optimization methods are implemented with Casadi \cite{anderssonCasADiSoftwareFramework2019} in Python.

\subsection{Optimality and Sample Efficiency}
\begin{figure}[h]
  \centering
  \includegraphics[width=1.0\linewidth]{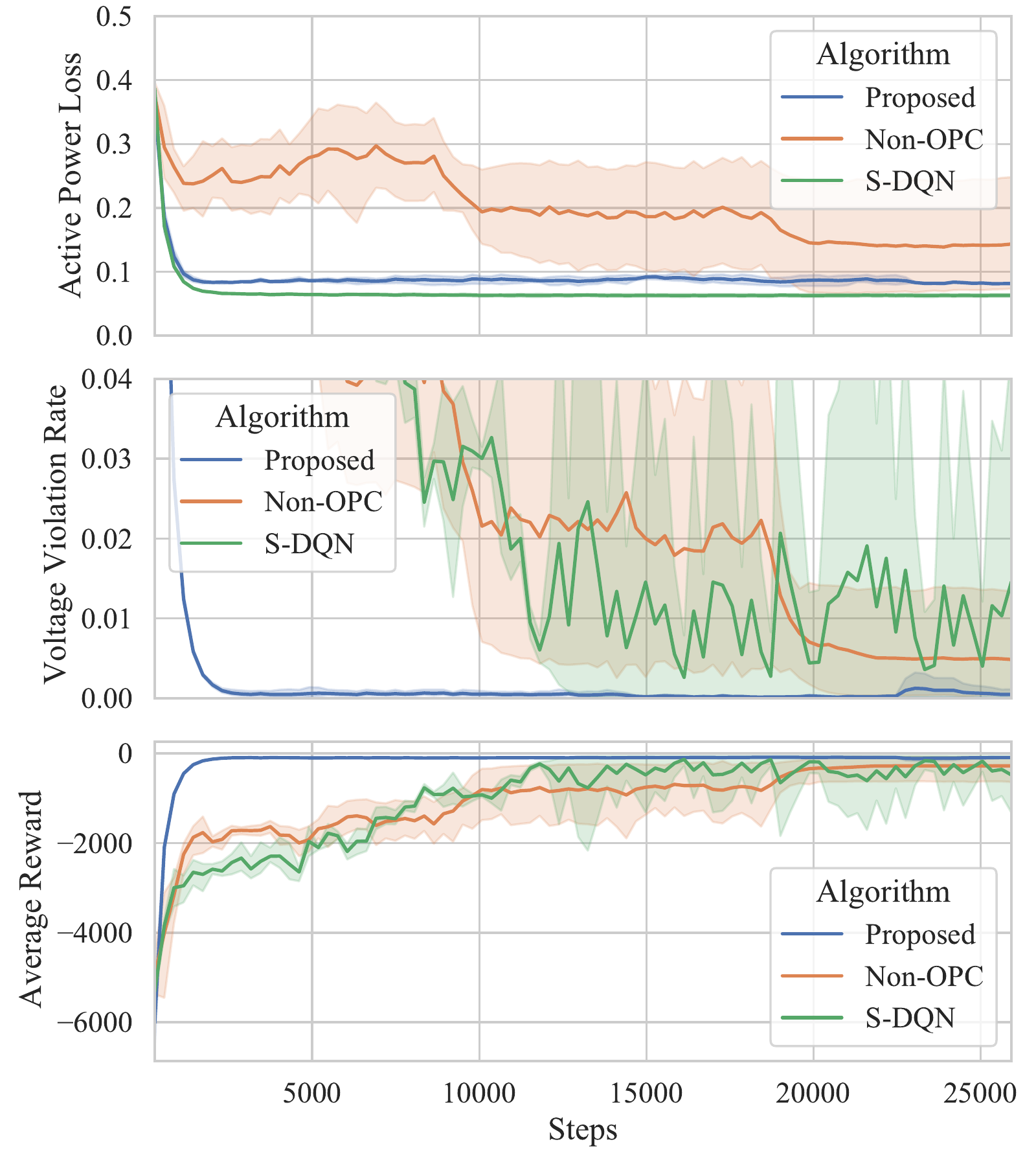}
  \caption{The training curve of DRL-based methods on the modified 33-bus ADN test feeder.}
  \label{fig:33bw}
\end{figure}

During the experiment, it is observed that 90 episodes are enough for the benchmarks and proposed method to get reasonable solutions.
The training process, which is supposed to evaluate on a real system in practice, is simulated and visualized in \cref{fig:33bw}.
Since DRL-based algorithms are stochastic methods, experiments are carried out with 3 independent random seeds, and the mean values and error bounds of performances are represented as solid curves and light-colored regions.

Comparing the proposed method with S-DQN, \cref{fig:33bw} illustrates that the proposed method achieves near-optimal performance on rewards with significantly fewer samples, even without any ADN topology or parameters information.
On the other hand, it is observed that S-DQN is better regarding active power loss from the very beginning.
This is because S-DQN takes advantage of oracle ADN models, which are not required by the proposed method, for FTA and achieves optimal reactive power distribution of DGs in every step.
But for the STA, the efficiency and stability of DQN is surpassed by the proposed MDSAC.

Comparing with A-OPT, the proposed method has avoided the misleading of errors in models as shown in \cref{tbl:final}.
Since the proposed method is data-driven and model-free, its performance may differ from the optimization-based one with oracle model.
The major advantage of our proposed method is to learn continuously online, coordinate two-timescale devices together and achieve near-optimal performance in a model-free manner.

\begin{table}[h]
  \centering
  \begin{threeparttable}
    \caption{Quantified indices of the trained agent and benchmarks in final episode}
    \label{tbl:final}
    \begin{tabular}{@{}lrrr@{}}
      \toprule
      \multicolumn{1}{r}{\multirow{2}{*}{Algorithm}} & \multicolumn{1}{c}{$P_\text{loss}/\text{MW}$}    & \multicolumn{1}{c}{VVR/p.u.}  & \multicolumn{1}{c}{$T_\text{loss}/\text{taps}$}    \\
      \cmidrule(l){2-4} & \hspace{30pt} 33-bus & \hspace{30pt} 33-bus & \hspace{30pt} 33-bus \\ \midrule
      Proposed     & 8.30e-02 & 0 & 2.4 \\
      S-DQN        & 6.05e-02 & 5.83e-02 & 4.3 \\
      A-OPT        & 1.08e-01 & 1.98e-01 & 7.0 \\
      OPT          & 4.59e-02 & 0   & 3.0 \\ \bottomrule
    \end{tabular}
  \end{threeparttable}
\end{table}

\subsection{Effectiveness of MTOPC}
As shown in \cref{fig:33bw}, our proposed method with MTOPC achieves near-optimal level of performance in the first 2000 steps, while the ``Non-OPC'' one without MTOPC needs much more samples and shows higher variance with different random seeds.
With certain seeds Non-OPC still learns satisfactory policies after iterations, but with other seeds it doesn't.
The reason is that the expectation itself we have corrected by MTOPC in \cref{eq:mtopc} is a random variable.
Even though the estimation is biased, the STA still learns in some cases.
Nevertheless, with MTOPC the bias can be eliminated and it improves the stability of the training process.
\begin{figure}[h]
  \centering
  \includegraphics[width=1.0\linewidth]{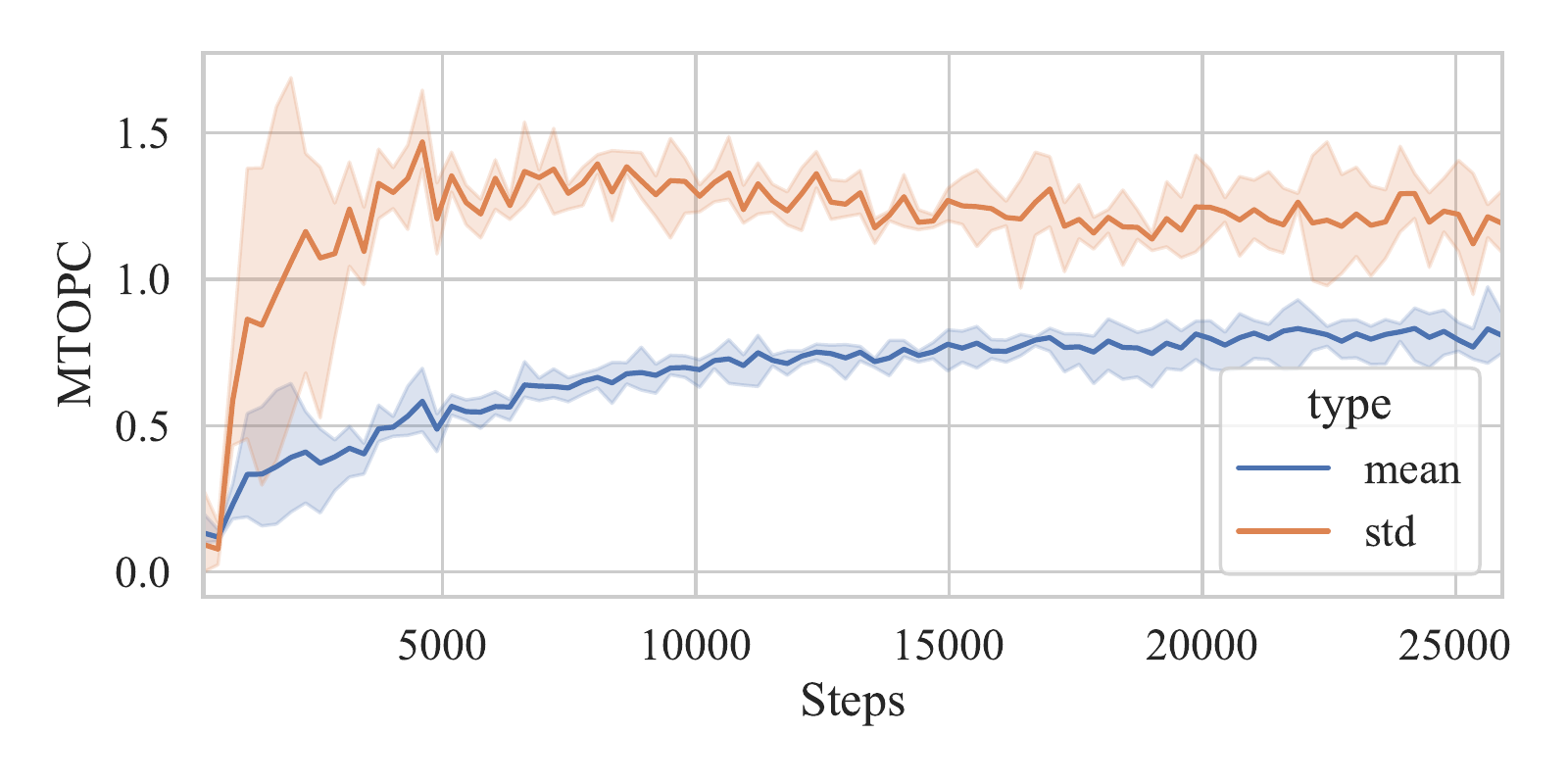}
  \caption{The mean and standard deviation of MTOPC's $\omega'$ on the modified 33-bus ADN test feeder with the proposed method.}
  \label{fig:opc33bw}
\end{figure}

To explain the effectiveness of MTOPC further, \cref{fig:opc33bw} the mean and standard deviation of MTOPC's $\omega'$ in a batch with the proposed method.
Similarly, the solid curves stand for mean value and light-colored regions stand for error bounds across random seeds.
In the starting periods, the mean $\omega'$ is near the lower bound because the policy of FTA is learning form the scretch.
But the deviation of $\omega'$ is at a relatively high level, which means the correction is carried out by reducing some samples and increasing the others.
While the FTA growing mature, the mean of $\omega'$ grows near 1 since the old samples still exist in $\mathcal{D}$. 
Also, it is worth mentioning that MTOPC not only benefits STA but also FTA because if the STA finds an optimal policy soon enough, FTA can save the effort of learning on the states with bad STA actions.

\section{Conclusion}
\label{sec:conclusion}
A bi-level off-policy DRL algorithm is proposed for two-timescale VVC with both FTCDs and STDDs coordinated without oracle network topology or parameters.
BMDP is defined to formulate the two-timescale VVC problem, and separate agents STA and FTA are implemented with off-policy DRL methods.
For STA, a high efficient off-policy algorithm MDSAC is proposed with value decomposition to address the curse of dimensionality.
For the coordination between STA and FTA, a novel MTOPC method is proposed to eliminate the inherent disturbance between the two agents.
Numberical studies conducted on the 33-bus distribution test feeders supports that the proposed algorithm achieves stable and satisfactory optimization of STDDs and FTCDs in a model free manner and outperforms existing two-timescale VVC.

\ifCLASSOPTIONcaptionsoff
  \newpage
\fi

\bibliographystyle{IEEEtran}
\bibliography{paper,phd}

\begin{thebibliography}{10}
\providecommand{\url}[1]{#1}
\csname url@samestyle\endcsname
\providecommand{\newblock}{\relax}
\providecommand{\bibinfo}[2]{#2}
\providecommand{\BIBentrySTDinterwordspacing}{\spaceskip=0pt\relax}
\providecommand{\BIBentryALTinterwordstretchfactor}{4}
\providecommand{\BIBentryALTinterwordspacing}{\spaceskip=\fontdimen2\font plus
\BIBentryALTinterwordstretchfactor\fontdimen3\font minus
  \fontdimen4\font\relax}
\providecommand{\BIBforeignlanguage}[2]{{%
\expandafter\ifx\csname l@#1\endcsname\relax
\typeout{** WARNING: IEEEtran.bst: No hyphenation pattern has been}%
\typeout{** loaded for the language `#1'. Using the pattern for}%
\typeout{** the default language instead.}%
\else
\language=\csname l@#1\endcsname
\fi
#2}}
\providecommand{\BIBdecl}{\relax}
\BIBdecl

\bibitem{kurbatovaGlobalTrendsRenewable2020}
T.~Kurbatova and T.~Perederii, ``Global trends in renewable energy
  development,'' in \emph{2020 {{IEEE KhPI Week}} on {{Advanced Technology}}
  ({{KhPIWeek}})}, Oct. 2020, pp. 260--263.

\bibitem{liuReactivePowerVoltage2009}
M.~B. Liu, C.~A. Canizares, and W.~Huang, ``Reactive {{Power}} and {{Voltage
  Control}} in {{Distribution Systems With Limited Switching Operations}},''
  \emph{IEEE Transactions on Power Systems}, vol.~24, no.~2, pp. 889--899, May
  2009.

\bibitem{borghettiUsingMixedInteger2013}
A.~Borghetti, ``Using mixed integer programming for the volt/var optimization
  in distribution feeders,'' \emph{Electric Power Systems Research}, vol.~98,
  pp. 39--50, 2013.

\bibitem{liuDistributedVoltageControl2018}
H.~J. Liu, W.~Shi, and H.~Zhu, ``Distributed {{Voltage Control}} in
  {{Distribution Networks}}: {{Online}} and {{Robust Implementations}},''
  \emph{IEEE Transactions on Smart Grid}, vol.~9, no.~6, pp. 6106--6117, Nov.
  2018.

\bibitem{xuAcceleratedADMMBasedFully2020}
T.~Xu and W.~Wu, ``\BIBforeignlanguage{en}{Accelerated {{ADMM}}-{{Based Fully
  Distributed Inverter}}-{{Based Volt}}/{{Var Control Strategy}} for {{Active
  Distribution Networks}}},'' \emph{\BIBforeignlanguage{en}{IEEE Transactions
  on Industrial Informatics}}, vol.~16, no.~12, pp. 7532--7543, Dec. 2020.

\bibitem{zhuFastLocalVoltage2016}
H.~Zhu and H.~J. Liu, ``Fast {{Local Voltage Control Under Limited Reactive
  Power}}: {{Optimality}} and {{Stability Analysis}},'' \emph{IEEE Transactions
  on Power Systems}, vol.~31, no.~5, pp. 3794--3803, Sep. 2016.

\bibitem{liuOnlineMultiagentReinforcement2021}
H.~Liu and W.~Wu, ``Online {{Multi}}-agent {{Reinforcement Learning}} for
  {{Decentralized Inverter}}-based {{Volt}}-{{VAR Control}},'' \emph{IEEE
  Transactions on Smart Grid}, pp. 1--1, 2021.

\bibitem{arnoldModelFreeOptimalControl2016}
D.~B. Arnold, M.~{Negrete-Pincetic}, M.~D. Sankur, D.~M. Auslander, and D.~S.
  Callaway, ``Model-{{Free Optimal Control}} of {{VAR Resources}} in
  {{Distribution Systems}}: {{An Extremum Seeking Approach}},'' \emph{IEEE
  Transactions on Power Systems}, vol.~31, no.~5, pp. 3583--3593, Sep. 2016.

\bibitem{wangSafeOffpolicyDeep2019}
W.~Wang, N.~Yu, Y.~Gao, and J.~Shi, ``Safe {{Off}}-policy {{Deep Reinforcement
  Learning Algorithm}} for {{Volt}}-{{VAR Control}} in {{Power Distribution
  Systems}},'' \emph{IEEE Transactions on Smart Grid}, pp. 1--1, 2019.

\bibitem{gaoBatchConstrainedReinforcementLearning2020}
Y.~Gao, W.~Wang, J.~Shi, and N.~Yu, ``Batch-{{Constrained Reinforcement
  Learning}} for {{Dynamic Distribution Network Reconfiguration}},'' \emph{IEEE
  Transactions on Smart Grid}, pp. 1--1, 2020.

\bibitem{nachumDataEfficientHierarchicalReinforcement2018}
O.~Nachum, S.~Gu, H.~Lee, and S.~Levine, ``Data-{{Efficient Hierarchical
  Reinforcement Learning}},'' \emph{arXiv:1805.08296 [cs, stat]}, Oct. 2018.

\bibitem{ecoffetFirstReturnThen2021}
A.~Ecoffet, J.~Huizinga, J.~Lehman, K.~O. Stanley, and J.~Clune,
  ``\BIBforeignlanguage{en}{First return, then explore},''
  \emph{\BIBforeignlanguage{en}{Nature}}, vol. 590, no. 7847, pp. 580--586,
  Feb. 2021.

\bibitem{lazaridisDeepReinforcementLearning2020}
A.~Lazaridis, A.~Fachantidis, and I.~Vlahavas, ``\BIBforeignlanguage{en}{Deep
  {{Reinforcement Learning}}: {{A State}}-of-the-{{Art Walkthrough}}},''
  \emph{\BIBforeignlanguage{en}{Journal of Artificial Intelligence Research}},
  vol.~69, pp. 1421--1471, Dec. 2020.

\bibitem{serbanArtificialIntelligenceSmart2020}
A.~C. {\c S}erban and M.~D. Lytras, ``Artificial {{Intelligence}} for {{Smart
  Renewable Energy Sector}} in {{Europe}}\textemdash{{Smart Energy
  Infrastructures}} for {{Next Generation Smart Cities}},'' \emph{IEEE Access},
  vol.~8, pp. 77\,364--77\,377, 2020.

\bibitem{stanojevReinforcementLearningApproach2020}
O.~Stanojev, O.~Kundacina, U.~Markovic, E.~Vrettos, P.~Aristidou, and G.~Hug,
  ``A {{Reinforcement Learning Approach}} for {{Fast Frequency Control}} in
  {{Low}}-{{Inertia Power Systems}},'' \emph{arXiv:2007.05474 [cs, eess]}, Jul.
  2020.

\bibitem{zhangResearchAGCPerformance2020}
D.~Zhang, H.~Zhang, X.~Zhang, X.~Li, K.~Ren, Y.~Zhang, and Y.~Guo, ``Research
  on {{AGC Performance During Wind Power Ramping Based}} on {{Deep
  Reinforcement Learning}},'' \emph{IEEE Access}, vol.~8, pp.
  107\,409--107\,418, 2020.

\bibitem{liuTwostageDeepReinforcement2020}
H.~Liu and W.~Wu, ``Two-stage {{Deep Reinforcement Learning}} for
  {{Inverter}}-based {{Volt}}-{{VAR Control}} in {{Active Distribution
  Networks}},'' \emph{IEEE Transactions on Smart Grid}, pp. 1--1, 2020.

\bibitem{liCoordinationPVSmart2019}
C.~Li, C.~Jin, and R.~Sharma, ``Coordination of {{PV Smart Inverters Using Deep
  Reinforcement Learning}} for {{Grid Voltage Regulation}},'' in \emph{18th
  {{IEEE International Conference}} on {{Machine Learning}} and
  {{Applications}} - {{ICMLA}} 2019}, 2019.

\bibitem{caoMultiAgentDeepReinforcement2020}
D.~Cao, W.~Hu, J.~Zhao, Q.~Huang, Z.~Chen, and F.~Blaabjerg, ``A
  {{Multi}}-{{Agent Deep Reinforcement Learning Based Voltage Regulation Using
  Coordinated PV Inverters}},'' \emph{IEEE Transactions on Power Systems},
  vol.~35, no.~5, pp. 4120--4123, Sep. 2020.

\bibitem{yangTwoTimescaleVoltageControl2020}
Q.~Yang, G.~Wang, A.~Sadeghi, G.~B. Giannakis, and J.~Sun, ``Two-{{Timescale
  Voltage Control}} in {{Distribution Grids Using Deep Reinforcement
  Learning}},'' \emph{IEEE Transactions on Smart Grid}, vol.~11, no.~3, pp.
  2313--2323, May 2020.

\bibitem{chenReinforcementLearningDecisionMaking}
X.~Chen, G.~Qu, Y.~Tang, S.~Low, and N.~Li,
  ``\BIBforeignlanguage{en}{Reinforcement {{Learning}} for
  {{Decision}}-{{Making}} and {{Control}} in {{Power Systems}}: {{Tutorial}},
  {{Review}}, and {{Vision}}},'' p.~16, 2021.

\bibitem{zhengRobustReactivePower2017}
W.~Zheng, W.~Wu, B.~Zhang, and Y.~Wang, ``Robust reactive power optimisation
  and voltage control method for active distribution networks via dual
  time-scale coordination,'' \emph{IET Generation, Transmission Distribution},
  vol.~11, no.~6, pp. 1461--1471, 2017.

\bibitem{xuMultiTimescaleCoordinatedVoltage2017}
Y.~Xu, Z.~Y. Dong, R.~Zhang, and D.~J. Hill, ``Multi-{{Timescale Coordinated
  Voltage}}/{{Var Control}} of {{High Renewable}}-{{Penetrated Distribution
  Systems}},'' \emph{IEEE Transactions on Power Systems}, vol.~32, no.~6, pp.
  4398--4408, Nov. 2017.

\bibitem{jinTwoTimescaleMultiObjectiveCoordinated2019}
D.~Jin, H.~Chiang, and P.~Li, ``Two-{{Timescale Multi}}-{{Objective Coordinated
  Volt}}/{{Var Optimization}} for {{Active Distribution Networks}},''
  \emph{IEEE Transactions on Power Systems}, vol.~34, no.~6, pp. 4418--4428,
  Nov. 2019.

\bibitem{jhaBiLevelVoltVAROptimization2019}
R.~R. Jha, A.~Dubey, C.~Liu, and K.~P. Schneider, ``Bi-{{Level Volt}}-{{VAR
  Optimization}} to {{Coordinate Smart Inverters With Voltage Control
  Devices}},'' \emph{IEEE Transactions on Power Systems}, vol.~34, no.~3, pp.
  1801--1813, May 2019.

\bibitem{zafarMultiTimescaleVoltageStabilityConstrained2020}
R.~Zafar, J.~Ravishankar, J.~E. Fletcher, and H.~R. Pota, ``Multi-{{Timescale
  Voltage Stability}}-{{Constrained Volt}}/{{VAR Optimization With Battery
  Storage System}} in {{Distribution Grids}},'' \emph{IEEE Transactions on
  Sustainable Energy}, vol.~11, no.~2, pp. 868--878, Apr. 2020.

\bibitem{loweMultiAgentActorCriticMixed2020}
R.~Lowe, Y.~Wu, A.~Tamar, J.~Harb, P.~Abbeel, and I.~Mordatch,
  ``\BIBforeignlanguage{en}{Multi-{{Agent Actor}}-{{Critic}} for {{Mixed
  Cooperative}}-{{Competitive Environments}}},''
  \emph{\BIBforeignlanguage{en}{arXiv:1706.02275 [cs]}}, Mar. 2020.

\bibitem{guQPropSampleEfficientPolicy2017}
S.~Gu, T.~Lillicrap, Z.~Ghahramani, R.~E. Turner, and S.~Levine, ``Q-{{Prop}}:
  {{Sample}}-{{Efficient Policy Gradient}} with {{An Off}}-{{Policy Critic}},''
  \emph{arXiv:1611.02247 [cs]}, Feb. 2017.

\bibitem{suttonMDPsSemiMDPsFramework1999}
R.~S. Sutton, D.~Precup, and S.~Singh, ``\BIBforeignlanguage{en}{Between
  {{MDPs}} and semi-{{MDPs}}: {{A}} framework for temporal abstraction in
  reinforcement learning},'' \emph{\BIBforeignlanguage{en}{Artificial
  Intelligence}}, vol. 112, no.~1, pp. 181--211, Aug. 1999.

\bibitem{haarnojaSoftActorcriticAlgorithms2018}
T.~Haarnoja, A.~Zhou, K.~Hartikainen, G.~Tucker, S.~Ha, J.~Tan, V.~Kumar,
  H.~Zhu, A.~Gupta, P.~Abbeel \emph{et~al.}, ``Soft actor-critic algorithms and
  applications,'' \emph{arXiv preprint arXiv:1812.05905}, 2018.

\bibitem{sunehagValueDecompositionNetworksCooperative2017}
P.~Sunehag, G.~Lever, A.~Gruslys, W.~M. Czarnecki, V.~Zambaldi, M.~Jaderberg,
  M.~Lanctot, N.~Sonnerat, J.~Z. Leibo, K.~Tuyls, and T.~Graepel,
  ``\BIBforeignlanguage{en}{Value-{{Decomposition Networks For Cooperative
  Multi}}-{{Agent Learning}}},'' \emph{\BIBforeignlanguage{en}{arXiv:1706.05296
  [cs]}}, Jun. 2017.

\bibitem{wangDuelingNetworkArchitectures2016}
Z.~Wang, T.~Schaul, M.~Hessel, H.~{van Hasselt}, M.~Lanctot, and N.~{de
  Freitas}, ``\BIBforeignlanguage{en}{Dueling {{Network Architectures}} for
  {{Deep Reinforcement Learning}}},''
  \emph{\BIBforeignlanguage{en}{arXiv:1511.06581 [cs]}}, Apr. 2016.

\bibitem{mnihAsynchronousMethodsDeep2016}
V.~Mnih, A.~P. Badia, M.~Mirza, A.~Graves, T.~P. Lillicrap, T.~Harley,
  D.~Silver, and K.~Kavukcuoglu, \emph{Asynchronous {{Methods}} for {{Deep
  Reinforcement Learning}}}, 2016.

\bibitem{schulmanProximalPolicyOptimization2017}
J.~Schulman, F.~Wolski, P.~Dhariwal, A.~Radford, and O.~Klimov,
  ``\BIBforeignlanguage{en}{Proximal {{Policy Optimization Algorithms}}},''
  \emph{\BIBforeignlanguage{en}{arXiv:1707.06347 [cs]}}, Aug. 2017.

\bibitem{brockmanOpenAIGym2016}
G.~Brockman, V.~Cheung, L.~Pettersson, J.~Schneider, J.~Schulman, J.~Tang, and
  W.~Zaremba, \emph{{{OpenAI Gym}}}, 2016.

\bibitem{baranNetworkReconfigurationDistribution1989}
M.~E. Baran and F.~F. Wu, ``Network reconfiguration in distribution systems for
  loss reduction and load balancing,'' \emph{IEEE Transactions on Power
  Delivery}, vol.~4, no.~2, pp. 1401--1407, 1989.

\bibitem{anderssonCasADiSoftwareFramework2019}
J.~A.~E. Andersson, J.~Gillis, G.~Horn, J.~B. Rawlings, and M.~Diehl,
  ``{{CasADi}} \textendash{} {{A}} software framework for nonlinear
  optimization and optimal control,'' \emph{Mathematical Programming
  Computation}, vol.~11, no.~1, pp. 1--36, 2019.

\end{thebibliography}
\end{document}